\documentstyle[aps,epsf,psfig,rotate,multicol]{revtex}
\begin{document}

\title{Application of a renormalization group algorithm to nonequilibrium
cellular  automata with one absorbing state}
\author{Roberto A. Monetti}
\address{Institut f\"ur Theoretische Physik \\ Physik-Department der 
Technischen Universit\"at M\"unchen \\  James-Franck-Strasse, 85747 Garching, 
Germany}
\author{Javier ~E. Satulovsky} 
\address{Department of Chemistry, Purdue University, \\
West Lafayette, IN 47907-1393, USA}

\date{\today}
\maketitle

\begin{abstract}
We improve a recently proposed dynamically driven renormalization group 
algorithm for cellular automata systems with one absorbing state, introducing
spatial correlations in the expression for the transition probabilities. We
implement the renormalization group scheme considering three different
approximations which take into account correlations in the stationary
probability distribution. The improved scheme is applied to a probabilistic 
cellular automaton already introduced in the literature. 
\end{abstract}

\begin{multicols}{2}
\section{Introduction}

In order to study cellular automata systems displaying a second order
irreversible phase transition to an absorbing state characterized by a scalar
order parameter, a dynamical renormalization group (RG) algorithm \cite{OliSat}
has been recently proposed. These cellular automata models are in the directed
percolation (DP) universality class. The method is based on a Dynamically
Driven  Renormalization Group (DDRG) scheme (for a recent review about DDRG
see  \cite{VesZapLor}), which has been  successfully applied  to self-organized
critical phenomena, as sandpile  models \cite{VesZapPie} and forest fire models
\cite{LorVesZap}.

The basic idea introduced in \cite{LorVesZap} is to couple a real space  RG
scheme to a stationary condition that drives the RG group equations through the
parameter space. The stationary equations, involving the stationary
distribution, have to be approximated, since the form of the stationary
probability distribution is not known a priori, as in the case of systems in
equilibrium.  Oliveira and Satulovsky \cite{OliSat} showed, as proposed in
\cite{LorVesZap}, that results can be improved using more refined
approximations for the stationary probability  distribution.  The expression
for the transition probability used in \cite{OliSat} consists in a
product of independent one-site transition probabilities at every step of the
RG transformation.

In this work we exploit another aspect of  the scheme, not  considered before,
in order to include  additional correlations. In fact, correlations can be also
introduced in the  renormalization scheme if we allow the transition
probability to depend upon  more neighbors.  In the case of  non-equilibrium
models, the space in which the RG flows is the space spanned by the transition
probabilities. As we will see later, our approach broadens this space providing
more degrees of freedom to the RG trajectories that flow towards the fixed
points.

We apply the modified RG scheme to a probabilistic cellular 
automaton (PCA) with one absorbing state already introduced in the literature 
\cite{Bha}.  Using a block renormalization that properly treats the nature of
the absorbing state, we figure the value of the critical exponent of the
divergence of the spatial correlation length, $\nu _{\perp }$, using three
different approximations involving correlations among clusters of $1$, $2$, and
$4$ neighboring sites  in the lattice.  Our calculations for $\nu_{\perp }$, at
small orders of approximation in the mean field scheme for the stationary
distribution, give better values than the ones reported in reference
\cite{OliSat}.

The paper is organized as follows. We begin with a brief description of the
model proposed in \cite{Bha} and the general renormalization scheme.
After this, we define the algorithm used in this work and present the values
obtained for $\nu_{\perp }$.  Finally,  after mentioning the ideas involved 
in the simulation technique used to study the model, we present the values of
the whole set of critical exponents for the PCA obtained by means of dynamical
numerical simulations and stationary simulations. Our results are in well
agreement with the values corresponding to ($1+1$) DP, and differ considerably
from the ones reported in reference  \cite{Bha}.

\section{The Model}

The model studied in \cite{Bha} is a one-dimensional cellular automaton 
in which each site can be either vacant, $\sigma_i = 0$, or occupied by a
particle, $\sigma_i = 1$. At each time step, the state of a given site will
depend only on its previous state and the previous state of its nearest
neighbors. The transition probability $T(\sigma
|\sigma ^{\prime })$ from state $\sigma ^{\prime }=(\sigma _1^{\prime
},\sigma _2^{\prime },\cdots,\sigma _L^{\prime })$ to state $\sigma =(\sigma
_1,\sigma _2,\cdots ,\sigma _L)$ will be given by the product 
\begin{equation}
\label{one-site}
T(\sigma |\sigma ^{\prime })=\prod_{i=1}^L\tau(\sigma _i|\sigma _{i-1}^{\prime
},\sigma _i^{\prime },\sigma _{i+1}^{\prime }),  \label{1}
\end{equation}
where $L$ is the number of sites and $\tau(\sigma _i|\sigma _{i-1}^{\prime
},\sigma _i^{\prime },\sigma _{i+1}^{\prime })$ is the one-site transition
probability given by the following rules 
\begin{equation}
\label{2}
\begin{array}{|c|c|c|c|c|c|c|c|c|} \hline
\tau & 000 & 001 & 100 & 101 & 010 & 011 & 110 & 111 \\ \hline
0 & 1 & 1\!\!-\!\!p & 1\!\!-\!\!p & 1 & 
0 & 1 & 1 & 1 \\ \hline
1 & 0 & p & p & 0 & 1 & 0 & 
0 & 0 \\ \hline
\end{array}
\end{equation}

This probabilistic cellular automaton models a transition from elementary rule
4 to elementary rule 22 (following Wolfram's nomenclature scheme) \cite{Wolf}.
The system has two critical points, one of them at $p_1 = 0$ and the other at
$p_2 \approx 0.75$. The critical point at $p_1 = 0$ will be shown to be 
trivial,
in contrast to the result obtained in reference \cite {Bha}. Consequently, 
we will apply the RG to study the non-trivial transition point.
Rule $\tau(0|000)=1$ implies that the vacuum state is indeed an absorbing 
state. 

\section{Renormalization Scheme}

The RG scheme proposed in \cite{LorVesZap} is a real space RG scheme 
\cite{MazNolVal} in which one renormalizes the transition probability, $T$. 
The RG flow takes place in the space of parameters defining $T$. A blocking
procedure transforms cells of $b$ sites into one site at the new scale. In
order to account for the fact that the vacuum state is absorbing, a cell
devoided of particles will always renormalize into an empty site. Cells with
at least one particle have been chosen to renormalize into an occupied site. 
Other options have been tried, but they do not preserve the existence of
the absorbing state.

Let $\sigma =(\sigma _1,\sigma _2,\cdots ,\sigma _L)$ be the state of a system
with $L$ degrees of freedom, and the vector  $S =(S_1,S_2, \cdots ,
S_{L^{\prime}})$ be the state of the renormalized system with $L^{\prime }=L/b$
degrees of freedom, where $b$ is the size of the renormalization block.

The conditional probability of state $S$ given state $\sigma$, 
${\cal R}(S |\sigma )$, must satisfy 
\begin{equation}
{\cal R}(S |\sigma )\geq 0,\qquad \qquad \sum_S {\cal R}(S |\sigma)=1.  
\label{4}
\end{equation}

Given $T$ and the probability of a state $\sigma ^{\prime }$ at time
$t$, $W(\sigma ^{\prime })$, one can write the joint probability of
state $\sigma ^{\prime }$ at time $t$ and state $\sigma $ at $n$ time
steps later, $W_n(\sigma ,\sigma ^{\prime })$, by simply applying $T$ to 
$W(\sigma ^{\prime })$ $n$ successive times 

\begin{equation}
W_n(\sigma ,\sigma ^{\prime })=T^n(\sigma |\sigma ^{\prime })
W(\sigma ^{\prime }).
\label{5}
\end{equation}
In addition, in the stationary regime, the probability distribution 
$W(\sigma )$ must satisfy

\begin{equation}
W(\sigma )=\sum_{\sigma ^{\prime }}T^n(\sigma |\sigma ^{\prime })
W(\sigma ^{\prime }),
\label{6}
\end{equation}
for any value of $n$.

In the same way one can write these expressions at the coarse grained level. 
Denoting by $\widetilde{T}(S ,S ^{\prime })$ 
the probability of occurrence of state $S^{\prime }$ 
at a given time and state $S$ one time step later, the RG
transformation is obtained imposing \cite{MazNolVal} 
\begin{equation}
\widetilde{W}(S ,S ^{\prime })=\sum_\sigma \sum_{\sigma ^{\prime }}
{\cal R}(S |\sigma ){\cal R}(S ^{\prime }|\sigma ^{\prime })W_n(\sigma
,\sigma ^{\prime }),  \label{7}
\end{equation}
from which follows 
\begin{equation}
\widetilde{W}(S ^{\prime })
=\sum_S \widetilde{W}(S ,S ^{\prime })
=\sum_{\sigma ^{\prime }}{\cal R}(S
^{\prime }|\sigma ^{\prime })W(\sigma ^{\prime }).  \label{8}
\end{equation}

Once knowing the transition probabilities at the coarse grained scale, one can 
easily build the rescaled transition matrix, 
$\widetilde{T}(S |S ^{\prime })$, as

\begin{equation}
\widetilde{T}(S |S ^{\prime })=\frac{\widetilde{W}(S ,S ^{\prime
})}{\widetilde{W}(S ^{\prime })}.  \label{10}
\end{equation}

Using now equations (\ref{5}), (\ref{7}), and (\ref{8}), we obtain the 
final expression for the  renormalization equations \cite{VesZapLor} 
\begin{equation}
\widetilde{T}(S |S ^{\prime })=\frac{\sum_\sigma \sum_{\sigma
^{\prime }}{\cal R}(S |\sigma ){\cal R}(S ^{\prime }|\sigma ^{\prime
})T^n(\sigma |\sigma ^{\prime })W(\sigma ^{\prime })}
{\sum_{\sigma ^{\prime }}{\cal R}
(S ^{\prime }|\sigma ^{\prime })W(\sigma ^{\prime })}.  
\label{11}
\end{equation}

Let us note that, while this equation expresses rescaled transition 
probabilities, $\widetilde{T}$, in terms of transition probabilities at a 
lower
scale, $T$, the stationary  weight of each state present in equation
(\ref{11}),   $W(\sigma^{\prime })$,  is yet unknown. Contrary to the case of
closed systems in thermal equilibrium, we do not know a priori the expression
for the stationary probability distribution. However,  including equation
(\ref{6}), one can get a closed set of equations to solve at  each
renormalization step. The stationarity condition (\ref{6}) is actually 
essential in driving the RG equations (\ref{11}) through parameter space.
Equation (\ref{11}), together with  a given approximation for the stationary
probability,  provide then a well defined RG transformation  $T \rightarrow
\widetilde{T}$.

In practice, equation (\ref{6}) can hardly be solved and one must resort to
approximations.  The values of critical exponents obtained using  the present
RG approach are expected to improve as  these approximations improve. We
have used three different levels  of approximations, in which
correlations among clusters of $1$, $2$, and $4$ sites are considered
respectively. 

In this work we will be concerned with another way to improve the  RG scheme.
An important point is that, in order to solve equations (\ref{11}), an
assumption needs to be made on how the transition probability between states $S
^{\prime }$ and $S$, $\widetilde{T}(S |S ^{\prime })$, depends on local 
transition probabilities at the coarse grained level. This choice will 
determine the degree of proliferation that the RG will have, since the form of
the renormalized transition probability will be preserved along the RG
trajectories.

In the former approach \cite{OliSat}, the authors carried out the most general 
RG transformation considering one-site transition probabilities. By preserving
the form of the renormalized transition probabilities, the RG trajectories are
found to flow to the attractive fixed points in a five-dimensional space
spanned by the  dynamical parameters. In contrast to usual RG methods in which
new couplings arise at each step of the transformation, this RG procedure is
not able to proliferate the dynamical parameters since the form of the
transition probabilities are kept fixed at the coarse grained level. However,
new dynamical parameters can be considered from the very beginning if we allow
the transition probabilities to depend upon more neighbors, i.e., introducing
more correlations in them. This feature of the method should be compared  with
other dynamical RG procedures \cite{MazVal} where the introduction of new
couplings since the very beginning is an alternative way to carry out the RG
transformation.

We propose, then, a form for the coarse grained transition probability
consisting of a product of independent two-site transition  probabilities
instead of one-site transition probabilities.  Denoting the position of each
lattice site with $i$, the new  transition probability is defined at even time
steps as 
\begin{equation}
\widetilde T(S |S ^{\prime})=
\!\!\!\prod_{k=1,(i=2k)}^{L/2}\!\!\!\widetilde \tau(S _i, S _{i+1}|
S _{i-1}^{\prime},S _i^{\prime},S _{i+1}^{\prime},S _{i+2}^{\prime}),
\label{12.even}
\end{equation} 
and at odd time steps as
\begin{equation}
\widetilde T(S |S ^{\prime})=
\!\!\!\!\!\prod_{k=1,(i=2k+1)}^{L/2}\!\!\!\!\!\!\!\!
\widetilde \tau(S _i,S _{i+1}|
S _{i-1}^{\prime},S _i^{\prime},S _{i+1}^{\prime},S _{i+2}^{\prime}).
\label{12.odd}
\end{equation} 
Here we have used the same symbol, $\widetilde \tau$, to indicate a different
type of transition probability than the ones appearing in (\ref{one-site}). In
formulas (\ref{12.even})  and (\ref{12.odd}) periodic boundary conditions are
assumed. 

One can retrieve at any time one-site transition probabilities  knowing both
two-site probabilities and the stationary distribution. It is straightforward
to show that 
\begin{eqnarray}
\widetilde \tau(S _i|&&S _{i-1}^{\prime},S _i^{\prime},S _{i+1}^{\prime})
= \!\!\!\!\!\!\!\!\sum_{S _{i+1},S _{i+2}^{\prime}} \!\!\!\!\!\!\!
\widetilde \tau(S _i,S _{i+1}|S _{i-1}^{\prime},S _i^{\prime},S 
_{i+1}^{\prime},
S _{i+2}^{\prime}) \nonumber \\ 
&&\frac{W(S _{i-1}^{\prime},
S^{\prime} _i,S _{i+1}^{\prime},S _{i+2}^{\prime})}{W(S _{i-1}^{\prime},
S^{\prime} _i,S _{i+1}^{\prime})}.
\end{eqnarray}

Using expressions (\ref{12.even}) and (\ref{12.odd}) in equation (\ref{11}) 
one can implement the RG transformation. The algorithm we used is explained in
more detail in the next section.

\section{Renormalization algorithm}

We have used a temporal coarse graining of two time steps ($n=2$). The
blocking  operator, ${\cal R}$, was chosen in the same way as in \cite{OliSat},
renormalizing cells of size $b=2$ into one site
\begin{equation}
{\cal R}(S |\sigma )=\prod_{k=1}^{L/2}R(S _k|\sigma _{2k-1},\sigma_{2k}),  
\label{21}
\end{equation}
with 
\begin{equation}
R(S _k|\sigma _{2k-1},\sigma _{2k})\geq 0, 
\end{equation}
and\begin{equation}
\sum_{S_k}R(S _k|\sigma _{2k-1},\sigma _{2k})=1.  \label{22}
\end{equation}
In order to preserve the 
nature of the absorbing state, we have also required $R$ to satisfy
\begin{equation}
R(0|0,0)=1,  \label{23}
\end{equation}
and 
\begin{equation}
R(0|\sigma _{2k-1},\sigma _{2k})=0.  \label{24}
\end{equation}
whenever $\sigma _{2k-1}\neq 0,$ or $\sigma _{2k}\neq 0$

\begin{figure}
\narrowtext
\centerline{\rotate[r]{\epsfysize=3.3in \epsffile{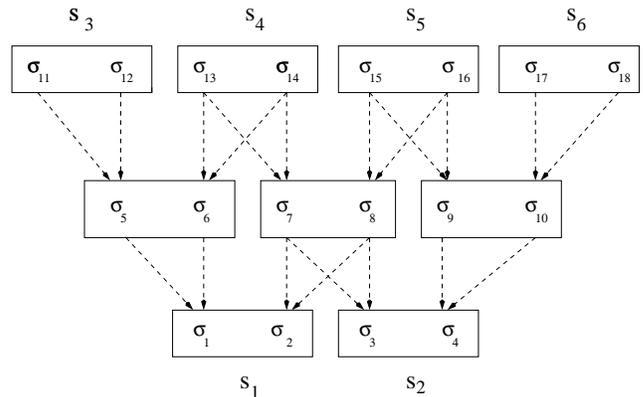}}}
\caption{Diagram showing the blocking scheme procedure. Numbers
correspond to the indexes used in equation~(\ref{31})-(\ref{33}).}
\label{scheme}
\end{figure}

The diagram in figure (\ref{scheme}) indicates how two-site transition
probabilities are renormalized. Indexes appearing in
equation ~(\ref{31})-(\ref{33}) refer to this diagram.

Using equations (\ref{11}), (\ref{12.even}), and (\ref{12.odd}), we can write
down the expression relating $\tau$ to $\widetilde{\tau}$, which is given by
\begin{eqnarray}
&&\widetilde{\tau}(S _1,S_2,|S_3,S _4,S _5,S_6)=
{[N(S_3,S_4,S_5,S_6)]}^{-1} \nonumber \\ 
&&
\sum_{\sigma _1 \sigma _2 \sigma _3 \sigma _4 \sigma _{11} \cdots 
\sigma _{18}}\!\!\!\!\!\!\!\!\!\!\!\!\! 
R(S _1|\sigma _1,\sigma _2) R(S _2|\sigma _3,\sigma _4) 
R(S_3|\sigma _{11},\sigma _{12})  \nonumber \\
&&
\;\;\;\;\;\;\;\;
R(S_4|\sigma _{13},\sigma _{14})
R(S_5|\sigma _{15},\sigma _{16})
R(S_6|\sigma _{17},\sigma _{18}) \nonumber \\ 
&&
D(\sigma _1,\sigma _2,\sigma _3,\sigma _4|\sigma _{11},\cdots ,\sigma _{18}) 
W(\sigma _{11},\cdots,\sigma _{18}),
\label{31}
\end{eqnarray}
where 

\begin{eqnarray}
&&D(\sigma _1\sigma _2,\sigma _3,\sigma _4 |\sigma _{11},\sigma _{12},
\sigma _{13},\sigma _{14},\sigma_{15},\sigma _{16},\sigma _{17},\sigma _{18})
= \nonumber \\
&&\sum_{\sigma _5 \cdots \sigma _{10}}
\widetilde{\tau}(\sigma _1,\sigma _2|\sigma _5,\sigma _6,\sigma _7,\sigma _8)
\widetilde{\tau}(\sigma _3,\sigma _4|\sigma _7,\sigma _8,\sigma _9,\sigma 
_{10})
 \nonumber \\
&&
\widetilde{\tau}(\sigma _5,\sigma _6|\sigma_{11},\sigma_{12},\sigma_{13},\sigma
 _{14})
\widetilde{\tau}(\sigma _7,\sigma _8|\sigma _{13},\sigma _{14},\sigma _{15},
\sigma _{16}) \nonumber \\  
&&
\widetilde{\tau}(\sigma _9,\sigma _{10}|\sigma _{15},\sigma _{16},\sigma _{17},
\sigma_{18})
^{\mbox{\ }},  \label{32}
\end{eqnarray}
and 
\begin{eqnarray}
N&&(S_3,S_4,S_5,S_6)= 
\sum_{\sigma _{11} \cdots \sigma _{18}}\!\!\!\!\! 
R(S_3|\sigma _{11},\sigma _{12}) R(S_4|\sigma_{13},\sigma _{14})\nonumber \\ 
&&R(S_5|\sigma _{15},\sigma _{16}) R(S_6|\sigma _{17},\sigma _{18}) 
W(\sigma _{11},\cdots,\sigma _{18}).
\label{33}
\end{eqnarray}

Since we do not know a priori the stationary weights, $W(\sigma_{11},
\cdots,\sigma _{18})$, we need an approximate method to estimate them. The
simplest approximation, sometimes known as simple mean field approximation,
consists in neglecting correlations among different sites. That is

\begin{equation}
W(\sigma _{11}, \cdots ,\sigma _{18})= 
\prod_{i=11}^{18}W(\sigma _i),  \label{41}
\end{equation}
where $W(\sigma _i)$ is the solution of 
\begin{eqnarray}
W(\sigma _1)=&&\sum_{\sigma _2\sigma _3\sigma _4\sigma _5\sigma _6}
\widetilde{\tau}(\sigma _1\sigma _2|\sigma _3\sigma_4\sigma _5\sigma _6) 
W(\sigma _2)\nonumber \\ 
&& W(\sigma _3)W(\sigma _4)W(\sigma _5)W(\sigma _6).  
\label{42}
\end{eqnarray}

Correlations, however, are actually taken into account in the geometrical
aspects of the blocking procedure, leading to non classical exponents.
Better approximations can be also implemented (as a reference see 
\cite{TaniaRon}). We used one, two, and four-site approximations.

Equations (\ref{31})-(\ref{33}) involve each transition probability. Being so
many terms, we are prevented from an analytical determination of the fixed 
points of the transformation. So, we performed our search numerically, using
initial values for the transition probabilities that correspond to the original
model (table (\ref{2})).

As a technical remark, let us say that in each  iteration of the RG, and given
a set of parameters, $\{\widetilde \tau\}$, we need to solve equation 
(\ref{42}) (or its analogue for two and four-site approximations) before the
next RG step. We have done this by iterating the equation until reaching
convergence.

\section{RG results}

\begin{figure}
\narrowtext
\centerline{{\epsfysize=2.7in \epsffile{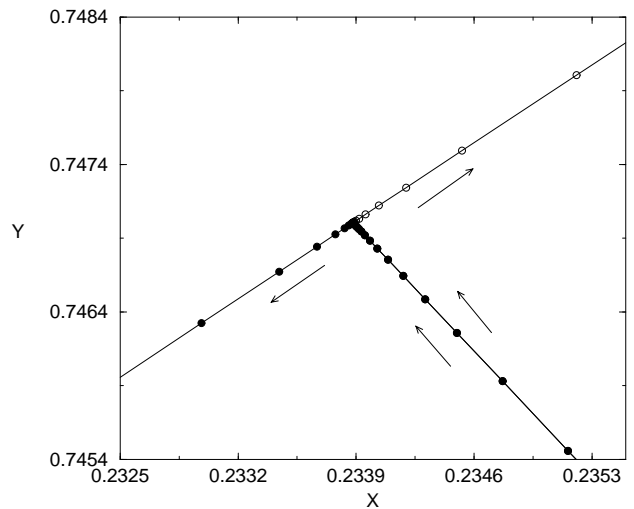}}}
\caption{Two dimensional projection of the RG flow along 
$x=\widetilde{\tau}(00|0010)$ and $y=\widetilde{\tau}(00|1000)$ 
for the simple mean field approximation.}
\label{RGflow}
\end{figure}

The behavior of the RG equations can be described as follows.  For values of
$p$ which are small enough, the set of transition probabilities flows  towards
an attractive fixed point characterized by a lattice devoid of particles. 
Increasing $p$ above a critical value, $p_{cr}$, the flow is driven to another
attractive fixed point, consisting of a lattice full of particles. The value of
$p_{cr}$, for each level of approximation used, can be found in table
(\ref{LambdaNu}). Starting around the critical values, the representative point
of the parameter set spends a long time near an unstable fixed point before
leaving towards one of the two attractive fixed points. A projection of the RG
flow along two specific transition probabilities is shown in figure
(\ref{RGflow}).

We have found one relevant parameter. Since we are only dealing with stationary
properties of the model, it is reasonable to assume that this parameter is
associated to the divergence of the spatial correlation length and not to the
temporal correlation length \cite{OliSat}. In order to calculate the
eigenvalue, $\Lambda$, associated to that parameter, we have to find the linear
region of the RG transformation. 

Let us take a trajectory passing close enough to the unstable point, and
construct a sequence of numbers consisting in the distance between two
successive points along that trajectory. Now, let us call $r$ to the ratio
between two consecutive numbers in that sequence. In the portion of the
trajectory corresponding to the linear region of the RG transformation (around
the unstable fixed point), one expects to see two plateaus  in the values of
$r$. 

\begin{figure}
\narrowtext
\centerline{{\epsfysize=2.5in \epsffile{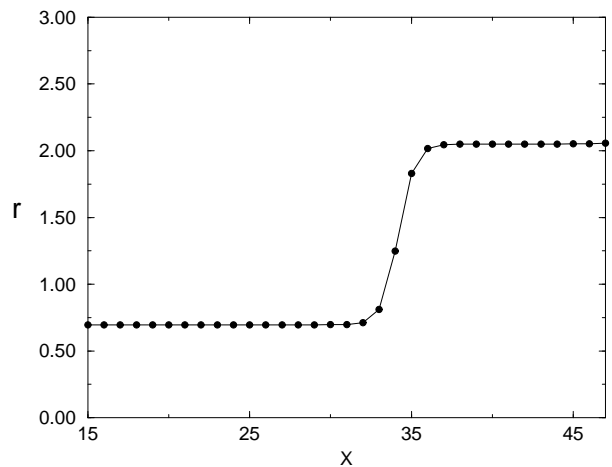}}}
\caption{The ratio, $r$, of successive distances among two consecutive points
for one of the trajectories shown in figure (\ref{RGflow}). The x coordinate 
denotes the precedence of each point along the curve.}
\label{ratios}
\end{figure}

The first plateau corresponds to a trivial parameter, while the value of
$r$ at the second plateau corresponds to the eigenvalue $\Lambda$ of the RG
transformation.
Figure (\ref{ratios}) shows an example of one of such curves 
for the simple mean field approximation.
So, figuring the eigenvalue $\Lambda$ associated to the relevant 
parameter, we get  $\nu _{\perp }=\ln 2/\ln \Lambda $. The value measured
numerically for the plot in figure (\ref{ratios}) (simple mean field
approximation) is $\nu _{\perp }=0.965 \pm 0.001$.

To the best of our knowledge, the best value of $\nu _{\perp }$ is $\nu _{\perp
} = 1.0972 \pm 0.0005$ \cite{BunH}, and the most accurate value for the
critical point  $p_{2}$, obtained by dynamical numerical simulations (see
following sections), is $p_{2} \approx 0.7513$. Although our value of the
critical exponent  $\nu _{\perp }$ is still inaccurate, it is worth at this
point comparing it with the one obtained using the same approximation
(simple mean-field) in \cite{OliSat}. In that work the authors obtained $\nu
_{\perp } = 0.931 \pm 0.005$, so that our result is a better approximation for 
the actual $\nu_{\perp }$. 
Our scheme is able to take into account correlations in a
more accurate way than the original one.

By increasing the order of the approximation, results improve, as shown in
table (\ref{LambdaNu}).
The value we found using the two-site approximation was 
$\nu _{\perp }=1.013 \pm 0.001$, while the value found in 
the four-site approximation was $\nu _{\perp}=1.015 \pm 0.001$, 
which is closer to the actual one.
Below we show the
critical value, $p_{cr}$, for the three approximations, as well as the
corresponding value of $\Lambda$ and  $\nu_{\perp}$
\begin{equation}
\begin{array}{|c|c|c|c|} \hline
\text{appr.}\!& p_{cr}\! & \Lambda \!& \nu _{\perp }  \!  \\ \hline
\text{1} & 0.639825  &  2.050  & 0.963 \pm 0.001  \\ \hline
\text{2} & 0.681490  &  1.982  & 1.013 \pm 0.001  \\ \hline
\text{4} & 0.695017  &  1.979  &  1.015 \pm 0.001   \\ \hline
\end{array}
\label{LambdaNu}
\end{equation}

It should be noted that the present RG scheme leads to fairly good values
for the exponent $\nu _{\perp }$ already within lower order mean-field
approximations. This fact indicates that the introduction of new correlations
in the transition probabilities plays a relevant role within the lower order
mean-field approaches. For mean-field approximations of order higher than two,
the convergence of the scheme becomes slower.

Reconsidering the ideas which led us to Eqs. (\ref{12.even}) and (\ref{12.odd})
for the transition probabilities, one may think of obtaining better
approximations for $\nu _{\perp }$  by allowing the transition probabilities to
depend upon even more neighboring lattice sites.  While this idea is clearly
right, one can presently not overcome, in practice, the huge amount of computer
time needed to obtain values that are accurate enough.

\section{Simulation Technique}
\subsection{Spreading analysis}

The basic rules governing the dynamical evolution of the system have been
formulated in section II. As in ordinary cellular automata, all lattice sites
are updated simultaneously. Simulations were performed on lattices of size $L
= 10000$, taking periodic boundary conditions. We briefly discuss here the
scaling theory for directed percolation which supports the spreading
analysis. A detailed treatment can be found elsewhere \cite{GrasTor}.

It should be stressed that for $L$ finite, the steady state of the system is 
meta-stable since, due to fluctuations of the stochastic process, there is  
always a finite probability for the system to become empty. This probability 
increases when approaching the critical point. Consequently, it is very
difficult to calculate critical points and critical exponents by means of
numerical simulations. Furthermore, since the transition between the
stationary regime and the absorbing state is second order, a mean field
treatment is not adequate. These shortcomings can be avoided by evaluating
critical exponents related to the dynamic critical behavior of the system. For
this purpose one starts, at $t = 0$, with a particle at the center of the
lattice otherwise empty, i.e. a configuration very close to the absorbing
state. Then, the following quantities are computed: (i) the survival
probability, $P(t)$, that is, the probability that at least a particle is still
in the system at time  $t$; (ii) the average number of particles, $N(t)$, and;
(iii) the average mean distance, $R(t)$, over which particles have spread.
Averages are taken over $5\times 10^4$ samples, and runs are performed up to $t
= 10^4$. Finite size effects are  avoided, since the epidemic disk never
reaches the edge of the lattice during the simulation. Close to the critical
point, and for long enough times, the following scaling laws should  hold 
\cite{GrasTor}:

\begin{equation}
P(t) \propto t^{- \delta} \Phi\{ \Delta \; t^{1/\nu_\|} \},
\end{equation}

\begin{equation}
N(t) \propto t^{\eta} \phi\{ \Delta \; t^{1/\nu_\|} \},
\end{equation}

\begin{equation}
R(t) \propto t^{z/2} \Xi\{ \Delta \; t^{1/\nu_\|} \},
\end{equation}
where $\Delta = |p - p_c|$, $\xi_t =  \Delta^{- \nu_\|}$ gives
the temporal correlation length close to $p_c$, $\nu_\|$ is the correlation 
length exponent (time direction), $\Phi$, $\phi$, and $\Xi$ are suitable 
scaling functions, and $\delta$, $\eta$, and $z$ are critical exponents. 
In the absorbing state, $P(t)$ and $N(t)$ are expected to decay exponentially,
since correlations are short-ranged. This can only happen if $\phi(\Delta,t)
\propto \{ \Delta \; t^{1/\nu_\|} \}^{- \eta \nu_\|} \exp(- \Delta^{\nu_\|}
t)$ for $t \rightarrow \infty$. Therefore, one has from Eq. (4):

\begin{equation}
\label{short_range}
N(t) \propto \Delta^{- \eta \nu_\|} \exp(- \Delta^{\nu_\|} t), \hskip 0.5 
true cm  t \rightarrow \infty.
\end{equation}

At criticality, one expects that log-log plots of $P(t)$, $N(t)$, and $R(t)$
would give straight lines, while upward and downward deviations would occur
even slightly off-criticality. This behavior would allow a precise
determination of the critical point and the critical exponents $\delta$, 
$\eta$,
and $z$. It should be noted that by means of Eq. (\ref{short_range}) 
it would be also possible to calculate $\nu_\|$.

\subsection{Finite size scaling analysis}
As in standard second order phase transitions it is assumed that in the
supercritical region, and close to the
critical point, the system displays spatial correlations characterized by a
typical length scale, $\xi_s$, which diverges at criticality according to

\begin{equation}
\xi_s \propto \Delta^{- \nu_\bot}, \hskip 0.5 true cm \Delta
\rightarrow 0,
\end{equation}
where $\nu_\bot$ is the correlation length exponent in the spatial direction.
The natural order parameter of the model is the density of particles, $\rho$,
which at criticality depends on the system size $L$ and $\Delta$ as
\begin{equation}
\label{universal}
\rho(p,L) = L^{-\beta/\nu_\bot} f(\Delta L^{1/\nu_\bot}),
\end{equation}
where $f$ is a suitable scaling function and $\beta$ is the order parameter
critical exponent. For small positive $\Delta$, and $L \rightarrow \infty$,
$f(x)$ should have the following form
\begin{equation}
f(x) \propto x^{\beta},
\end{equation}
in order to recover the well known critical behavior of the order parameter in
the thermodynamic limit
\begin{equation}
\rho \propto \Delta^{\beta}.
\label{beta}
\end{equation}

\section{Simulation results}
Before presenting the simulation results we will briefly discuss the critical
point at $p = 0$. If we start at $t = 0$ with a random initial configuration of
density $\rho_{0} = 0.5$, the stationary density of the
system is $\rho \approx 1/8$. It should be noted that at $p = 0$ the system
reaches a static stationary state in one time step (see the evolution rules in 
table (\ref{2})). Taking into account this observation, the stationary density
of the system can be obtained as follows. It is clear from the evolution rules
that the probability of having an occupied site at $t = 1$ is equal to the 
probability of finding an occupied site surrounded by empty sites at $t = 0$. 
Since there are no correlations in the initial state, the stationary density 
of the system at $p = 0$ can be determined as
\begin{equation}
\rho = \rho_{0} (1 - \rho_{0})^2
\end{equation}
Then, since we used $\rho_{0} = 0.5$, the expected value of the stationary 
density is  $\rho = 1/8$.

For arbitrary small values of $p$, the stationary state is the vacuum state
\cite{Bha}.  We now consider the relaxation process to the vacuum state. We
take as  starting configuration ($t = 0$) any of the static stationary  states
at $p = 0$. Then, we follow the evolution of the density for $p$ close  to
zero. It is clear from the evolution rules that after a particle is created,
two particles  are removed from the system at the next time step. Then, the
system stays in  another static  state until the next creation process occurs.
Suppose that a creation process happens at position $x_i$. If no new particle
is created in the  neighborhood of $x_i$ at the next time step, correlations
can not be generated (see the evolution rules in table (\ref{2})). The
probability of two  consecutive creation processes is $p^2 < p$. Then, the
system can not develop long range correlations and a mean field analysis should
be appropriate. For $p$ close to zero it is possible to consider the creation
process as a creation-induced  annihilation process. We then have
\begin{equation}
d \rho / dt = -p\rho.
\end{equation}
Consequently, $p = 0$ is a trivial
critical point since the relaxation time behaves as $\tau = {(1/p)}^1$. This
result is in disagreement with the one reported in reference \cite{Bha},
probably due to a poor statistics of the simulation data.

In the following, the results of the epidemics analysis at $p = p_2$ are
presented.
We measure the time evolution of $P(t)$, $N(t)$, and $R(t)$ for  
different values of the parameter $p$. Log-log plots of these quantities as a
function of time are straight lines at the phase transition and show curvature
away from the transition. 
It is important to mention that the epidemics analysis is a very sensitive 
method since it is
possible to distinguish among supercritical and sub-critical behavior for $p$
values that differ in the fourth decimal. Our best estimation of the critical
point is $p_2 = 0.7513 \pm 0.0002$, and the dynamical critical exponents are
$\delta = 0.162 \pm 0.0004$, $\eta = 0.304 \pm 0.0005$, and $z/2 = 0.643 \pm 
0.0007$. It should be remarked that the error bars merely indicate the
statistical error obtained from regressions. The values of the dynamical 
exponents are in well agreement with those
corresponding to directed percolation in ($1+1$) dimensions, as it was 
expected. 

It is possible to calculate the exponent $\nu_\|$ from the analysis of the 
sub-critical behavior (see eq. (\ref{short_range})). 
In fact, the decay constant, $\lambda = 
\xi_t^{-1}$, governing the long time behavior of $N(t)$, behaves according to
\begin{equation}
\lambda = \xi_t^{-1} = \Delta^{\nu_\|},
\end{equation}
so, if $p_2$ is known, we can calculate $\lambda$ for different values of
$\Delta$. Then, a log-log plot of $\lambda$ vs $\Delta$ allows us to evaluate
the exponent $\nu_\|$. This analysis gives an exponent $\nu_\| = 1.738 \pm
0.002$ which is quite close to the value $\nu_\| = 1.73$ corresponding to 
(1+1) DP \cite{BunH}.  It should be pointed out
that our value of $\nu_\|$ sharply differs from the one reported in reference
\cite {Bha} ($\nu_\| \approx 1.087$). The error in the last value of the 
exponent $\nu_\|$ is due to the fact that it was calculated taking into 
account not only sub-critical, but also supercritical curves.

We have also calculated the order parameter critical exponent measuring  the
density, $\rho$, as a function of  $\Delta$ in the supercritical regime  (see
eq.(\ref{beta})). We obtain $\beta = 0.277 \pm 0.002$, which is once again very
close to $\beta = 199/720$, corresponding to ($1+1$) DP \cite{BunH}.  It should
be mentioned that the reported value of the order parameter critical exponent
in reference  \cite{Bha} is $\beta \approx 0.32$ which differs around $16 \%$
from the theoretical value. This difference may be due again to the poor
statistics of the data.

\begin{figure}
\narrowtext
\centerline{\rotate[r]{\epsfysize=3.3in \epsffile{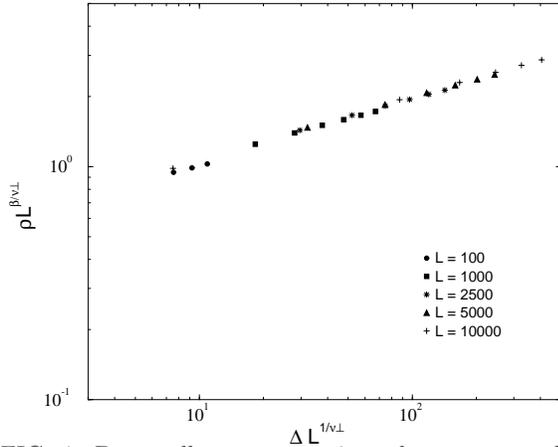}}}
\caption{Data collapse on a universal curve, according to equation
(\ref{universal}).}
\label{collapse}
\end{figure}

We finally present the finite size scaling analysis. Figure (\ref{collapse}) 
shows a log-log plot of $\rho L^{\beta/\nu_{\bot}}$ vs $\Delta
L^{1/\nu_{\bot}}$ for different values of $p$, and lattice sizes, $L$, where we
have used  $\beta=199/720$ and $\nu_{\bot} = 1.0972$ \cite {BunH} corresponding
to ($1+1$) DP. We obtain an excellent collapse of the data on an universal
curve, as it is predicted  by eq. (\ref{universal}).  

Right before submitting this manuscript for publication, the author of
reference \cite{Bha} improved some of his previous results \cite{Bha1}. 
Although no new results are reported
for the trivial critical point $p_1$, the conclusions concerning the
universality class of the model are in complete agreement with the ones 
found in this work.

\section{Conclusions}

We have introduced a new renormalization group algorithm for probabilistic   
cellular automata with one absorbing state. The new scheme introduces
correlations in the RG procedure by allowing the transition probabilities to 
depend upon two neighboring lattice sites. Three different approximations for
the stationary probability distribution have been used, namely the simple 
mean-field approximation, the pair mean-field approximation, and the 
four-site mean-field approximation. 

The new RG scheme leads to fairly good
values for $\nu_\bot$ even within mean-field approximations of low order. This
result shows that the introduction of spatial correlations in the transition
probabilities is the relevant reason for the improvement of the results.   

The critical exponents $\nu_\bot$, specially for low order approximations, are
better than the ones obtained with schemes that make use of an independent
product of one-site transition probabilities \cite{OliSat,TaMa}.

Using very simple arguments, we have shown that $p_1 = 0$ is a trivial critical
point, since the time relaxation constant behaves as $\tau = (1/p)^1$. This 
behavior differs from the one reported in reference \cite{Bha} ($\tau =
(1/p)^{0.86}$). 

We have also obtained, by means of numerical simulations, the critical point $p
= p_2$, and the whole set of critical exponents. The value of the critical
point  $p_2 = 0.7513 \pm 0.0002$ is in agreement with  \cite{Bha}. However, we
found very different values for exponents $\nu_{\|}$ and $\beta$. Our
values of the exponents at $p = p_2$ are in well agreement with those
corresponding to ($1+1$) DP as it was expected, since there is only one
absorbing state for the system \cite{Grass,Jann}. 

\section{acknowledgments}
R. ~A. ~M. would like to acknowledge CONICET (Argentina) for the provision of 
a fellowship. 


\end{multicols}
\end{document}